\documentclass[12pt]{article}
\usepackage{latexsym}
\newcommand{\be}{\begin{equation}}
\newcommand{\ee}{\end{equation}}
\newcommand{\bq}{\begin{eqnarray}}
\newcommand{\eq}{\end{eqnarray}}
\newcommand{\nalp}{n_{\alpha}}
\newcommand{\nbet}{n_{\beta}}
\newcommand{\npee}{n_{p}}
\newcommand{\nalpt}{n_{\tilde{\alpha}}}
\newcommand{\nbett}{n_{\tilde{\beta}}}
\begin{document}
\begin{titlepage}
\today          \hfill 
\begin{center}
\hfill    LBNL-43108 \\
          \hfill    UCB-PTH-99/20 \\

\vskip .5in

{\large \bf An Operator Approach\\
To String Equations}
\footnote{This work was supported by the Director, Office of Energy 
Research, Office of High Energy and Nuclear Physics, Division of High 
Energy Physics of the U.S. Department of Energy under Contract 
DE-AC03-76SF00098 and in part by the National Science Foundation under
grant PHY-95-14797.}

\vskip .5in
Korkut Bardakci\footnote{e-mail:kbardakci@lbl.gov}
\vskip .5in

{\em Department of Physics\\
University of California at Berkeley\\
   and\\
 Theoretical Physics Group\\
    Lawrence Berkeley National Laboratory\\
      University of California\\
    Berkeley, California 94720}
\end{center}

\vskip .5in

\begin{abstract}

In this paper, a new approach to string dynamics is proposed. String
coordinates are identified with a non-commuting set of operators
familiar from free string quantization, and the dyanamics follows
from the Virasoro algebra. There is a very large gauge group
operating on the non-commuting coordinates. The gauge has to be fixed
suitably to make contact with the standard string picture.

\end{abstract}
\end{titlepage}

\newpage
\renewcommand{\thepage}{\arabic{page}}
\setcounter{page}{1}
\noindent{\bf 1. Introduction}
\vskip 9pt
In the last couple of years, it has become clear that string theory
has a very rich structure, which was not apparent in its early
formulation [1]. There has been progress in several parallel and
overlapping directions: Various string theories which were once considered
distinct are now known to be related by duality transformations [2], and
new extended objects called branes were shown to play an important role
in string dynamics [3,4]. There is also the interesting new idea of
duality relating string theory in an AdS background to the conformal
field theory on the boundary of the AdS space [5,6]. In view of all of
these developments, there is clearly a need for a new approach 
 at a fundamental level which can unify and explain all of
these diverse aspects of string theory. Recently, such a unifying
approach called M(atrix) theory has been proposed [7,8], and it has
already led to anumber of interesting results.

A natural idea that has been around for some time is to attribute
the rich structure of string theory to an as yet undiscovered
gigantic gauge symmetry. From this point of view, the gauge symmetries
observed in the zero mass sector, the diffeomorphism invariance associated
with the graviton, as well as various dualities are the manifestations
of this hidden symmetry. However, the full structure of this
conjectured gauge symmetry is yet to be discovered.
 One reason why this symmetry is so difficult
to unmask in the usual perturbative formulation as well as in the
string field theory approach [9-12] is that in these formulations,
 there is always a good deal of gauge fixing to start with. In most
versions of string field theory, the gauge fixing is done either 
through the BRST method or by adopting the light cone gauge [13].
Both the original matrix theory and its compactified version, the 
 matrix string theory [14,15] operate again
 within the gauge fixed lightcone framework. Still missing
is  a formulation in terms of a large number of mostly
redundant coordinates and a big gauge group. In such a
formulation, various
different descriptions of string theory will correspond
to different choices of gauge.

This paper is an attempt to answer two questions:
 What is this big gauge gauge group and also what is the
space it acts upon ?
 The sigma model[16-20] description of the zero mass
modes of the string should shed some light on this problem. Imposing
conformal invariance on this model results in equations of motion which are
explicitly invariant under
 the gauge symmetries associated with the massless vector mesons and
coordinate transformations associated with gravity.
 However, the neglect of the heavy modes of the string makes it
impossible to go beyond these well known symmetries of the low energy
field theory limit. There is a version [21,22] of the sigma model which
incorporates the heavy modes of the string, and
 conformal invariance in this case is realized through 
  the Wilson renormalization group  approach. Unfortunately,
 the gauge symmetries of the
resulting equations are quite obscure, probably because of the particular
cutoff used. In a modified version [23] of this approach, invariance under
a certain class of coordinate transformations can be made explicit, but
this  only scratches the surface: A large number of symmetries are
still hidden.

The basic idea of this paper is to formulate string dynamics in such a
way that coordinates and momenta appear on equal footing. By coordinates
and momenta we mean, not just the zero modes, but the full set of
quantized modes of, for example, the free string. We will
essentially make use of a phase space description, with a doubling
of the coordinates. Corresponding to this enlarged coordinate space,
there will an enlarged gauge group, which will now include transformations
that mix non-commuting variables. This generalization is motivated by
two observations: Various generalized duality transformations [24]
interchange coordinates and momenta, and if they are to be part of the
proposed gauge group, transformations that mix coordinates and momenta
should be included. In addition, the emergence of non-commutative
geometry [25-31] as a viable description of some aspects of string
theory has been very suggestive. In a sense, what we have here is a
generalization of non-commutative geometry to include not just the 
zero modes, but all of the other modes of the string.

We need to specify a dynamical principle as the basis of our approach.
Usually, the starting point is an action, which is required to be 
conformally invariant. The stress tensor of this action then generates
a conformal algebra. We reverse this sequence of steps and start
with the conformal algebra as the basic principle. The generators
of the conformal algebra are assumed to be expressible in terms of the
operators that generate the free string spectrum, and the dynamical
equations we are seeking follow the requirement that the algebra close.
An important bonus of this approach is that
invariance under a big gauge group that mixes non-commuting variables 
is naturally built into the dynamical framework.

The paper is organized as follows: In section 2, we will first set up
the problem in general terms, and then discuss the perturbative
expansion around the flat (free string) background. As usual, such
a perturbative expansion seems the only way of making progress in
solving the dynamics. This enables one to linearize the equations
as a first approximation, and the rest of the section will mostly
deal with these linear equations. We will present some special
 solutions to them
 based on the vertex construction, and we will then introduce
a representation of the operators in terms of states, which will turn
out to be useful for a general analysis of the equations. In section 3,
we will study the open string dynamics in some detail, with the idea
of making contact with the usual free string equations. We will be able
to identify a state, which, after some gauge fixing, can be made to
depend only on a set of commuting coordinates, which can be identified
with position coordinates of the string. Moreover, this state satisfies
the standard free string equations of motion. In the next section,
these results are generalized to the closed string. In section 5,
we will go beyond the linear approximation and
turn our attention to the interaction term. A convenient expression
for this term will be given in terms of coherent state representation.
Finally, in section 6, the basic equations will be reformulated
using the BRST formalism.

\noindent {\bf 2. The General Setup}
\vskip 9pt

As explained in the introduction, our goal is to derive the dyanamics
of the bosonic string by constructing representations of the Virasoro algebra
\be
[L_{m},L_{n}]=(m-n)L_{m+n}+c(m^{3}-m)\delta_{m+n},
\ee
where c is the central charge. In the case of the closed string, we have to 
work with two commuting copies of this algebra. Any conformally invariant
two dimensional theory will supply a realization of this algebra; the best
known special case is the free field realization in terms of the operators
$\alpha_{m}^{\mu}$,
\be
L^{(0)}_{m}=\frac{1}{2}\sum_{-\infty}^{\infty}\alpha^{\mu}_{m-n}\alpha^{\mu}
_{n}.
\ee
The $\alpha$'s  satisfy the commutation relations
$$
[\alpha_{m}^{\mu},\alpha_{n}^{\nu}]=m\,\delta_{m+n}\eta^{\mu\nu}.
$$
We wish to generalize the free field realization to an interacting one, still
using the same set of operators. The most general expression for the
$L_{m}$'s can somewhat symbolically be written in the form
\be
L_{m}=\sum_{i}\phi(x)_{m}^{(i)}(:\prod \alpha:)_{i}.
\ee
In this equation, x is the position coordinate and the $\phi$'s are the
wavefunctions corresponding to various levels of the string. The sum goes
over all of the possible products of $\alpha$'s,
 normal ordered to avoid possible singular expressions. Imposing
the commutation relations of eq.(1) on this ansatz will result in a set
of equations for the $\phi$'s, which we can identify as string field equations.
These equations are easy to write down; but
solving them is another matter. The only practical method seems to be
a perturbation expansion around a background, which is an explicitly
known solution of the same algebra. Therefore, although the equations
are background independent, the perturbative solution will depend on the
background. In this paper, we will expand the L's around the free field
solution of eq.(2), corresponding to a flat background.
 Setting
$$
L_{m}=L_{m}^{(0)}+K_{m},
$$
gives
\be
[L_{m}^{(0)},K_{n}]-[L_{n}^{(0)},K_{m}]+[K_{m},K_{n}]
-(m-n)K_{m+n}=0.
\ee
Here we have assumed that the central charges of $L^{(0)}$'s and $L$'s
are the same.

In the next section, we will investigate systematically the linear
 part of this 
equation, dropping the last term quadratic in the K's:
\be
[L_{m}^{(0)},K_{n}]-[L_{n}^{(0)},K_{m}]-(m-n)K_{m+n}=0.
\ee
Eq.(5) is similar to the equation of motion of a non-abelian Chern-Simons
 gauge theory,
with the difference that ordinary commuting coordinate derivatives are
replaced by the non-commuting operators $L_{m}^{(0)}$. An important property
it shares with that theory
 is invariance under the  transformations
\be
L_{m}\rightarrow U^{-1}L_{m}U,
\ee
where U is an arbitrary unitary
 operator constructed out of $x$ and the $\alpha$'s.
These transformations can be identified with gauge transformations
under which the dynamics is invariant. They will play an important role
in what follows.

The linearized equations are  invariant under the abelian part of the gauge
transformations:
\be
K_{m}\rightarrow K_{m}+[L_{m},H].
\ee
As usual, solutions related by gauge transformations will be considered
equivalent.

In the case of closed strings, in addition to $L^{(0)}$'s, we have
$\tilde{L}^{(0)}$'s, which have an  expansion analogous to eq.(2)
 in terms of the $\tilde{\alpha}$'s, and there is also the corresponding
$\tilde{K}$'s. So in addition to eq.(4), there is an identical equation
for the variables with tildes, and also an equation that expresses the
condition that $L$'s commute with $\tilde{L}$'s:
\be
[L^{(0)}_{m},\tilde{K}_{n}]-[\tilde{L}^{(0)}_{n},K_{m}]
+[K_{m},\tilde{K}_{n}]=0.
\ee
It is easy to find special solutions to eq.(5) through the standard
vertex construction. The vertex operator, given by
\be
V(k,\tau)=:exp\left(ik.X(\tau)\right): 
\ee
where
$$
X^{\mu}(\tau)=x^{\mu}+\tau\,\alpha^{\mu}_{0}
 +i\sum_{1}^{\infty}\frac{1}{m}\left(
e^{-im\tau} \alpha^{\mu}_{m}- e^{i m\tau}\alpha^{\mu}_{-m}\right),
$$
satisfies the relation[ref]
\be
[L_{m}^{(0)},V(k,\tau)]=-i\,e^{im\tau}\frac{d}{d\tau}\left(V(k,\tau)\right)
+\frac{1}{2}m\, k^{2}e^{i m\tau}\,V(k,\tau).
\ee
As a result, we can set the K's equal to the vertex for tachyon emission:
\be
K_{m}=e^{i m\tau}\:V(k,\tau),
\ee
with the condition
$$
k^{2}=2,
$$
and eq.(5) is satisfied for any $\tau$ and any k that satisfies the
tachyon mass shell condition. Furthermore, this solution is abelian;
different K's commute, and as a result, it satisfies the initial
exact equation (4). We also note that different values of the parameter
$\tau$ correspond to gauge equivalent results, since
\be
\left[L_{m},\int_{\tau_{1}}^{\tau_{2}}d\tau\: V(k,\tau)\right]=
-i\left(e^{i m\tau_{2}}\:V(k,\tau_{2})-e^{i m\tau_{1}}\:V(k,\tau_{1})\right).
\ee
One can easily construct an infinite number of additional solutions for
the K's by replacing the tachyon emision vertex by the emission vertices
for the higher excited states.

The solution given above also generalizes to the closed string. In this case,
we have two commuting set of operators, $\alpha$'s and $\tilde{\alpha}$'s,
and two commuting set of Virasoro algebras generated by the $L$'s and
$\tilde{L}$'s. Let $\tilde{V}$ denote the vertex operator of eq.(8), 
constructed from $\tilde{\alpha}$'s instead of $\alpha$'s. Then, the
corresponding K's and $\tilde{K}$'s are given by
\bq
K_{m}&=&\int_{0}^{2\pi}d\sigma\: e^{i m(\tau-\sigma)}\,\tilde{V}(k,\tau-
\sigma) V(k,\tau+\sigma),\nonumber\\
\tilde{K}_{m}&=&\int_{0}^{2\pi}d\sigma\: e^{i m(\tau+\sigma)}\,
\tilde{V}(k,\tau-\sigma) V(k,\tau+\sigma),
\eq
with
$$
k^{2}=2.
$$
In this case, the solution  only satisfies the linearized
equations. Again, vertices for the emission of excited states provide
additional solutions. 

We now wish to analyze the linear equations for the K's in general,
using the operator expansion of eq.(3). We found it more convenient
to carry out this analysis in what amounts to a phase space representation
of the operators. In this representattion, a one to one
 correspondonce is set
up between the normal ordered operator products and a set of states.
The states are constructed by the following procedure: First,
a new set of operators, $\beta_{m}^{\mu}$, which commute with the
$\alpha$'s, are introduced, as well as a new c-number momentum P, which
commutes with x. Next,  the old non-commuting set of operators
are replaced by a new
set of commuting operators according to the recipe
$$
K_{m}\rightarrow |m\rangle, \;\;
\alpha_{-m}\rightarrow \alpha_{-m},\;\; \alpha_{m}\rightarrow \beta_{-m},
\;\; \alpha_{0}=-i\partial\rightarrow P,
$$
where $m>0$. Finally,
the application of the resulting operator on the vacuum gives the
state which is associated with the original operator.
 For example, the normal ordered operator product
$$
\phi(x)\,\alpha_{0}^{\mu}\,\alpha^{\nu}_{-m}\,\alpha^{\lambda}_{n}
$$
corresponds to the state
$$
\phi(x)\,P^{\mu}\,\alpha^{\nu}_{-m}\,\beta^{\lambda}_{-n}|0\rangle.
$$
According to our definition of the normal product, the operators
$\alpha_{0}^{\mu}=-i\partial^{\mu}$ are
 always placed to the right of $\phi(x)$.

We shall need the expressions for $L^{(0)}_{m}$'s acting on these states.
We start with a general normal ordered operator product, and we compute
its commutator with $L^{(0)}_{m}$ given by eq.(2). The original
operator product, as well as the final commutator, can be represented
by states, as explained above. We therefore have a mapping from one set
 of states to another induced by $L^{(0)}_{m}$'s, and this mapping
provides us with a representation different from eq.(2) for the
$L^{(0)}$'s. We will call it the state representation for the 
$L^{(0)}$'s and designate them by the same symbol as before when
no confusion can arise. It is straightforward to compute the explicit
expressions for $L^{(0)}$'s in this new representation; however,
as they stand, these expressions are a bit complicated. They can be
simplified considerably by making an additional canonical transformation
$$
\alpha_{m}\rightarrow \alpha_{m}-\beta_{-m},\;\;
\beta_{m}\rightarrow \beta_{m}-\alpha_{-m},
$$
where $m>0$. To avoid complicating the notation, we will use the same
letters for the new set of operators, it being understood that,
 from now on, $\alpha$'s and
$\beta$'s will stand for the new set of operators.
Finally, we have the following simple result:
\be
L^{(0)}_{m}\rightarrow L^{\alpha}_{m}-L^{\beta}_{-m},
\ee
where $L^{\alpha}$ is given by eq.(2) and $L^{\beta}$ has an identical
expansions in terms of the $\beta$'s. The only change is in the definition
of the zero modes; they are now given by
\be
\alpha^{\mu}_{0}=P^{\mu}-i\partial^{\mu},\;\;\;
\beta^{\mu}_{0}=P^{\mu}.
\ee
The lack of symmetry between the zero modes is due to our particular 
normal ordering prescription for the operators x and $-i\partial$
discussed earlier. We also note that the $L^{(0)}$'s defined by eq.(14)
have no central charge. This is to be expected, since the action of
$L^{(0)}$'s on the states is defined by their commutators with the
corresponding operators, and the central charge drops out of the picture
since it commutes with everything.

The linearized equations(eq.(5)), in this new language, are
\be
\left(L^{\alpha}_{m}-L^{\beta}_{-m}\right)|n\rangle -\left(L^{\alpha}_{n}
-L^{\beta}_{-n}\right)|m\rangle -(m-n)|m+n\rangle=0,
\ee
where the state $|m\rangle$ has replaced the operator $K_{m}$.
In the language of these new states,
 hermiticity properties of the original operator products
 are a bit complicated.
Hermitian conjugation of the operator product
 corresponds to the following interchange of the corresponding
states:
\be
\alpha_{m}\leftrightarrow \beta_{m},\;\; \phi\leftrightarrow 
\phi^{\ast},\;\;|m\rangle\leftrightarrow -|-m\rangle,\;\;P\leftrightarrow
P-i\partial.
\ee

\vskip 9pt
\noindent{\bf 3. Solutions to the Linear Equations for the Open
String}
\vskip 9pt

In this section, we will investigate the linearized equations for 
the open  string (eqn.(5) ), and connect the
solutions with the well-known states of the free bosonic string. 
 At first
sight, it seems that we have both too many states and 
 too many independent variables. Instead of a single string state,
there are infinitely many of them labeled by an integer m.
Also, the number of coordinates are doubled compared to the standard
string picture. The states are now constructed using two new sets
of operators: $\beta$'s in addition to
the $\alpha$'s, P in addition to x.
 On the other hand, there are also a  large set of gauge transformations
given by eq.(7), which translate into
\be
|m\rangle\rightarrow |m\rangle
+\left(L^{\alpha}_{m}-L^{\beta}_{-m}\right)|g\rangle,
\ee
where $|g\rangle$ represents an arbitrary ``gauge'' state. As we shall show, 
equivalence under these gauge transformations will drastically reduce
the number of independent states; only the state with $m=-1$ will turn
out to be independent. Also, this state will be a function of only
the $\alpha$'s and $x$, and not of the $\beta$'s and $P$.
This reduction from the original set of non-commuting  coordinates
to a final set of commuting  coordinates will be accomplished by a
(partial) gauge fixing.
 To see this, let us organize the states as follows:
Consider an expansion of the states
 in powers of P, and label each term in the expansion
 by $n_{p}$,
 the number of factors
of P it contains. In addition, we label them by the level numbers
$n_{\alpha}$ and $n_{\beta}$; they are respectively
 the eigenvalues of the operators
$$
N_{\alpha}=\sum_{1}^{\infty}\alpha_{-m}\alpha_{m},
$$
and
$$
N_{\beta}=\sum_{1}^{\infty}\beta_{-m}\beta_{m}.
$$
A given state $|m\rangle$ can therefore be written as a sum:
\be
|m\rangle=\sum_{n_{p},n_{\alpha},n_{\beta}}|m,n_{\alpha},n_{\beta},n_{p}
\rangle.
\ee
We note that all of the n's are non-negative integers. Also,
it is easy to show that the difference $n_{\alpha}-n_{\beta}$ is the
same for each term of this expansion.

We now focus on the state with $m=-1$. In principle, we could have started
with any other value of m, but it turns out that this particular value
of m has special properties that makes it easy to work with.
 We wish to establish the following
two results:\\
a) In the expansion of this state, all of the terms with either $n_{p}$ or
$n_{\beta}$ greater than zero can be gauged away.\\
b) After fixing gauge, the complete solution to eq.(16) can be expressed
in terms of  the single remaining component of $|-1\rangle$, namely
$|-1,n_{\alpha},0,0\rangle$. This state,
 which depends only x and the $\alpha$'s,
can be identified with usual string state.
 Furthermore, we will show that
it satisfies the correct mass shell equation and the subsidiary conditions.\\
Since we are contemplating an expansion in powers of P,
 it is convenient to seperate the L's into two parts;
one of them linear in P and the other P independent:
$$
L_{m}^{\alpha}= L_{m}^{\alpha}(0)+ P.\alpha_{m},\;\;\;
L_{m}^{\beta} = L_{m}^{\beta}(0)+ P.\beta_{m}.
$$
The condition for a state $|-1,\nalp,\nbet,\npee\rangle$ to be spurious (pure
gauge) follows from eq.(18). If gauge states $|g,\nalp,\nbet,\npee\rangle$,
satisfying the equation
\bq
&&L_{1}^{\beta}(0)|g,\nalp,\nbet+1,\npee\rangle= |-1,\nalp,\nbet,\npee\rangle
+L_{-1}^{\alpha}(0)|g,\nalp-1,\nbet,\npee\rangle\nonumber\\
&&+ P.\alpha_{-1}|g,\nalp-1,\nbet,\npee-1\rangle
+ P.\beta_{1}|g,\nalp, \nbet+1,\npee-1\rangle,
\eq
can be found, then this state is spurious. We wish to show that this equation
has solutions for states with either or both
 $\nbet$ and $\npee$ greater than or equal to 1. The basic idea is to solve
it by iteration, starting with the smallest values of $\nbet$ and $\npee$
and working up to larger values. The gauge states corresponding to the
smallest value of $\nbet$, $\nbet=0$, are not determined by the
above equation, and we  set them equal to zero for simplicity.
 Starting with $\nbet=0$, we note that this is a special case,
since the left hand side of eq.(20) vanishes:
\be
L_{1}^{\beta}(0)|\nbet=1\rangle=0.
\ee
In this case, we can solve for the gauge state $|g,\nalp,1,\npee\rangle$ by 
inverting the operator $P.\beta_{1}$:
\be
|g,\nalp,1,\npee\rangle= -\frac{1}{(\npee+1)!}
 \beta_{-1}.\frac{\partial}{\partial P}\left( 
|-1,\nalp,0,\npee+1\rangle \right).
\ee
This shows that all of the $m=-1$ states with $\nbet=0$ and $\npee\geq 1$
can be gauged away. Now let us consider the next set of states at
$\nbet=2$.
 In this case, we 
can write
\bq
&&|g,\nalp,\nbet+1,\npee\rangle=\left(L_{1}^{\beta}(0)\right)^{-1}\Big(
|-1,\nalp,\nbet,\npee\rangle
+ L_{-1}^{\alpha}(0)|g,\nalp-1,\nbet,\npee\rangle\nonumber\\
&&+ P.\alpha_{-1}|g,\nalp-1,\nbet,\npee-1\rangle
+ P.\beta_{1}|g,\nalp,\nbet+1,\npee-1\rangle\Big).
\eq
We note that the states on the right hand side of this equation all have
$\nbet\geq 1$, and acting on these states, the operator
 $L_{1}^{\beta}(0)$ is invertible. This is in contrast to the previous
case when, acting on states
with $\nbet=0$, $L_{1}^{\beta}(0)$ was not invertible, and that is why
eq.(20) for $\nbet=1$ had to be treated as a special case. Now taking
$\nbet=1$  and $\npee=0$ in eq.(23), the right hand side
of the equation is completely known. This is because the gauge states
with $\nbet=1$ are already known (eq.(22)), and the last term on the right
hand side vanishes since there is no state with $\npee=-1$. The left hand
side of this equation then gives us all of the gauge states with
$\nbet=2$ and $\npee=0$. Now take $\npee=1$ with again $\nbet=1$; again
the right hand side is completely known, and recycling the equation once more
gives us the gauge states with $\nbet=2$ and $\npee=1$. Continuing this
iteration, all of the gauge states with $\nbet=2$ and arbitrary $\npee$
are determined. Next we set $\nbet=2$ in eq.(23) and continue the double
iteration process till the gauge states for all of the values of
$\nbet$ and $\npee$ are determined. This argument establishes the result
that all of the states $|-1, \nalp, \nbet,\npee\rangle$ with either of or both
$\nbet$,$\npee$ greater than or equal to 1 can be gauged away. 

Having shown that the only surviving piece 
 of  the state with $m=-1$ has $\nbet=0$ and $\npee=0$, we now derive
 the equation satisfied by this state. First setting $n=-1$,
and  $m\geq 0$ in eq.(16),
and then projecting out the $\nbet=0$, $\npee=0$ component, we have
\bq
&&L_{m}^{\alpha}(0) |-1,\nalp+m,0,0\rangle
- L_{-1}^{\alpha}(0)|m,\nalp-1,0,0\rangle\nonumber\\
&&- (m+1)|m-1,\nalp,0,0\rangle =0.
\eq
Notice that $L_{-m}^{\beta}(0)$ does not contribute for $m\geq 0$ because
of the projection into $\nbet=0$ sector, and neither does $L_{1}^{\beta}(0)$ 
 because of the eq.(21). Consider the special case $m=0$ in
the above equation:
\be
\left(L_{0}^{\alpha}(0)-1\right)|-1,\nalp,0,0\rangle - L_{-1}^{\alpha}(0)
|0,\nalp-1,0,0\rangle=0.
\ee
We now argue that the state $|0,\nalp-1,0,0\rangle$ can be gauged away. Under
gauge transformations,
$$
|0,\nalp,0,0\rangle\rightarrow |0,\nalp,0,0\rangle
+ L_{0}^{\alpha}(0)|g,\nalp,0,0\rangle.
$$
The operator $L_{0}^{\alpha}(0)$ is always invertible, since
it contains a term of the form $\alpha_{0}^{2}=-\Box$, which can always
be inverted. Therefore, a suitable gauge state can always be chosen
to make the last term in eq.(25) vanish. With this gauge choice, we have the
standard mass shell condition:
\be
\left(L_{0}^{\alpha}(0)-1\right)|-1,\nalp,0,0\rangle =0.
\ee
It is important to notice that the choice $m=-1$ for the state in question
was crucial in arriving at the  mass shell condition given above.
With any other choice of m, there would be extra terms in the corresponding
equation; for example, the analogue of eq. (22) for
$m=-2$,
$$
L_{2}^{\beta}(0)|\nbet=2\rangle =0
$$
is simply not true.
It is possible to go beyond the mass shell condition and establish the standard
subsidiary conditions
\be
L_{m}^{\alpha}(0)|-1,\nalp,0,0 \rangle =0,\;\;\;m\geq 1,
\ee
 by showing that the states $|m,\nalp,0,0\rangle$
 for $m\geq 1$ can all be gauged
away. This can be done, for example,
 by an adaptation of the standard string theory
argument[1].

Having established that the state $|-1,\nalp,0,0\rangle $ can be made to
satisfy the standard string equations (eqs.(26) and (27)), the next step is
to show that, with further gauge fixing, all of the other states, up to
gauge equivalence, can
be determined in terms of this single state. In section (6), this will be
shown  using the BRST method; here, to illustrate how this process works
in a simple example, 
 we will construct an explicit solution
for the states with $\nbet=1$ and $\npee=0$ for $m\geq 0$.
 We take $n=-1$, $m\geq 0$
 in eq.(16), and project out the component
corresponding to $\nbet=0$ and $\npee=1$:
\bq
&&L_{m}^{\alpha}(0)|-1,\nalp+m,0,1\rangle - L_{-1}^{\alpha}(0)|m,\nalp-1,
0,1\rangle -(m+1) |m-1,\nalp,0,1\rangle\nonumber\\
&&+ P.\alpha_{m}\,|-1,\nalp+m,0,0\rangle
+ P.\beta_{1}\,|m,\nalp,1,0\rangle =0.
\eq
In this equation, the state $|-1,\nalp+m,0,1\rangle$ vanishes by the previous
gauge choice. The following ansatz provides a solution;
\bq
|m,\nalp,0,1\rangle&=&0,\;\;\;m\geq 0,\nonumber\\
|0,\nalp,1,0\rangle&=&i\beta_{-1}.\partial\,|-1,\nalp,0,0\rangle,\nonumber\\
|m,\nalp,1,0\rangle&=& -\beta_{-1}.\alpha_{m}|-1,\nalp+m,0,0\rangle
\;\;\;m\geq 1.
\eq
There is a consistency check: The solution given above should also satisfy
the $\nbet=1$,$\npee=0$ component of eq.(16) for $m,n\geq 0$,
\bq
&&L_{m}^{\alpha}(0)|n,\nalp+m,1,0\rangle - L_{n}^{\alpha}(0)
|m,\nalp+n,10\rangle -(m-n)|m+n,\nalp,1,0\rangle\nonumber\\
&&-\delta_{m,0}|n,\nalp,1,0\rangle +\delta_{n,0}|m,\nalp,1,0\rangle
=0,
\eq
and indeed it does. In principle, solutions for higher values of
$\nbet$ and $\npee$, as well as for the negative values of m can be
constructed in this fashion, although this brute force method soon
becomes too laborious.

We close this section with some further comments on gauge invariance.
Starting with a very rich gauge group (eq.(18)), a good deal of gauge
fixing had to be done to arrive at the string equations (26) and (27)
for $|-1, \nalp,0,0\rangle$, our candidate for the physical string
state. There is, however, still invariance under the residual
 gauge transformations
\be
|-1,\nalp,0,0\rangle \rightarrow |-1,\nalp,0,0\rangle
+ L_{-1}^{\alpha}(0)|g,\nalp-1,0,0\rangle,
\ee
where the gauge state satisfies
$$
L_{0}^{\alpha}(0)|g,\nalp-1,0,0\rangle =0.
$$
This is the well-known invariance under the shift by zero norm states
generated by $L_{-1}$, and it is valid in any dimension. What is
misssing is the invariance under shifts by zero norm states generated by
$L_{-m}$'s for $m>1$, which is only valid in the critical dimension. This
is a really troublesome puzzle, and we have only the following
tentative suggestion for a solution. It is possible that the choice
of the state with $m=-1$, and $\nbet=\npee=0$ for the physical string
state, although it has the advantage of simplifying various computations,
is not suitable for fully understanding the residual gauge invariance.
For example, the the choice of the state
\be
|s,\nalp,0,0\rangle = \sum_{m=-1}^{\nalp}|m,\nalp,0,0\rangle,
\ee
for the physical state has the advantage that, under gauge 
transformations,
\be
|s,\nalp,0,0\rangle\rightarrow |s,\nalp,0,0\rangle
+ \sum_{m=1}^{\nalp}L_{-m}^{\alpha}(0)|g,\nalp-m,0,0\rangle,
\ee
the gauge states are generated by all of the $L_{-m}$'s, not just by
$L_{-1}$. However, we have found states of this type difficult to
work with and we will not pursue this point any further.

\vskip 9pt
\noindent{\bf 4.Solutions to the Linear Equations for the Closed
String}
\vskip 9pt
In this section, we will briefly discuss the solutions to the linear
closed string equations. Our treatment will be less complete
compared to the case of the open string.
 The operators $K_{m}$ and $\tilde{K}_{m}$ are then
replaced by the states $|m\rangle$ and $|\tilde{m}\rangle$.
 There are  now two
copies of the equation (16), plus an equation expressing the
commutativity of the two Virasoro algebras (eq.(8)):
\bq
&&L_{m}^{(0)}|n\rangle -L_{n}^{(0)}|m\rangle -(m-n)|m+n\rangle=0,\\
&&\tilde{L}_{m}^{(0)}|\tilde{n}\rangle-\tilde{L}_{n}^{(0)}|\tilde{m}\rangle
-(m-n)|\tilde{m}+\tilde{n}\rangle=0,\\
&&\tilde{L}_{m}^{(0)}|n\rangle -L_{n}^{(0)}|\tilde{m}\rangle =0,
\eq
where,
$$
L_{m}^{(0)}=L_{m}^{\alpha}-L_{-m}^{\beta},\;\;
\tilde{L}_{m}^{(0)}= L_{m}^{\tilde{\alpha}}- L_{-m}^{\tilde{\beta}}.
$$
The gauge transformations are given by
\bq
|m\rangle&\rightarrow& |m\rangle+L_{m}|g\rangle,\nonumber\\
|\tilde{m}\rangle&\rightarrow&|\tilde{m}\rangle+\tilde{L}_{m}|g\rangle.
\eq
Our discussion of the closed string states will more brief and less complete
compared to the open string.
As in the last section, we expand the states in powers of P and
level numbers $\nalp$, $\nbet$, and now, in addition, in
level numbers $\nalpt$ and $\nbett$. The
states are thus labeled as
$$
|m,\nalp,\nbet,\nalpt,\nbett,\npee\rangle, \;\;\;
|\tilde{m},\nalp,\nbet,\nalpt,\nbett,\npee\rangle.
$$
The usual constraint between the right-left level numbers requires that
\be
\nalp-\nbet=\nalpt-\nbett.
\ee
We have now to make a choice for the physical string state. In parallel
with  the open string, we will focus on the states with
$m=-1$ and set
\be
|-1,\nalp,0,\nalpt,1,0\rangle= i\tilde{\beta}_{-1}.\partial\,
|s,\nalp,0,\nalpt,0,0\rangle,
\ee
where $\nalp=\nalpt$ and the state
$$
|s,\nalp,0,\nalpt,0,0\rangle
$$
is identified with the string state. There is, of course, some arbitrariness
in this choice, for example,
 the reason for having the operator $i\tilde{\beta}_{-1}.
\partial$ on the right is not yet clear. Also, this choice is not
symmetric between left and right moving (tilde and non-tilde) states.
 We hope to clarify these issues as
we proceed.

We have to show that the state defined above satisfies two sets of
equations; one set invoving the $L$'s and the other set $\tilde{L}$'s.
The first set of equations are easy to derive.
 Starting with eqn.(34), and following the
same steps as in the last section, one arrives at the analogues of
eqns.(26) and (27):
\bq
\left(L_{0}^{\alpha}(0)-1\right)|s,\nalp,0,\nalpt,0,0\rangle
&=& 0,\nonumber\\
L_{m}^{\alpha}(0)|s,\nalp,0,\nalpt,0,0\rangle &=&0,\;\;\; m\geq 1.
\eq
We now turn our attention to eqs.(35) and (36). In particular,
eq.(36) imposes
stringent conditions on the choice of the string state. Take
$m=n=-1$ and set   $\npee=0$ in this equation:
\be
\tilde{L}_{-1}^{(0)}(0)|m=-1,\npee=0\rangle=
L_{-1}^{(0)}(0)|\tilde{m}=-1,\npee=0\rangle.
\ee
For a non-trivial solution, both sides of this equation must vanish;
 otherwise, the operators $L_{-1}(0)$ and $\tilde{L}_{-1}(0)$
 would be
invertible and the resulting solution would be a pure gauge. We therefore
set
\be
\tilde{L}_{-1}(0)|-1,\npee=0\rangle = \left(L_{-1}^{\tilde{\alpha}}(0)
- L_{1}^{\tilde{\beta}}(0)\right)|-1,\npee=0\rangle=0.
\ee
To satisfy this condition, we write this state as a sum over $\nbett$,
$$
|-1,\npee=0\rangle=\sum_{\nbett=1}^{\infty}|-1,\nbett,\npee=0\rangle,
$$
and identify the term with $\nbett=1$ with the state defined by eq.(39).
We then have the following solution:
\be
|-1,\npee=0\rangle=\sum_{k=0}^{\infty}\left(L_{1}^{\tilde{\beta}}(0)
\right)^{-k}\left(L_{-1}^{\tilde{\alpha}}(0)\right)^{k}
(i\tilde{\beta}_{-1}.\partial)|s,\nalp,0,\nalpt,0,0\rangle.
\ee
We note that the operator $L_{1}^{\tilde{\beta}}(0)$ is invertible
 on states with $\nbett\geq 1$, and therefore each term on the right
hand side of the above equation is well defined. This is one of the
reasons for the presence of the operator $\tilde{\beta}_{-1}$ in the
definition of the state given by eq.(39); as a result, the sum in the
equation above starts at $\nbett=1$.

Next, we take $n=-1$ and $m\geq 0$ in eq.(36):
$$
\tilde{L}_{m}(0)|n=-1,\npee=0\rangle= L_{-1}(0)|\tilde{m}=m,\npee=0\rangle
\;\;m\geq 0.
$$
Again, unless both sides vanish, the candidate for the string state
would be a pure gauge. We must therefore have
$$
\left(L_{m}^{\tilde{\alpha}}(0)-L_{-m}^{\tilde{\beta}}(0)\right)
|-1,\npee=0\rangle=0,\;\;\;m\geq 0.
$$
Now project into the $\nbett=0$ sector in this equation. Because of this
projection, the operator $L_{-m}^{\tilde{\beta}}(0)$ does not contribute
for $m>0$, and at $m=0$, and acting on the state defined by eq.(39),
it gives  one. The resulting equations complement eqns.(40):
\bq
\left(L_{0}^{\tilde{\alpha}}(0)-1\right)|s,\nalp,0,\nalpt,0,0\rangle
&=&0,\nonumber\\
L_{m}^{\tilde{\alpha}}(0)|s,\nalp,0,\nalpt,0,0\rangle&=&0,
\;\;\;m\geq 1.
\eq
Another reason for having $\tilde{\beta}_{-1}$ in
eq.(39) is that it gives the correct mass shell condition in the
above equation.
\footnote{ Our normalization of the slope parameter for the closed
string differs by a factor of two from the conventional one.}

As an application of the formalism developed here, we note note that
T duality, under which the left moving(tilde) components are multiplied by 
-1, is easily implemented by a transformation of the form given by
eq.(6). Similarly, we hope that various other dualities can be
implemented by gauge transformations of the same type.
 An important point is that
although these symmetries are usually quite transparent in the original
non-gauge fixed form of the string equations, they are difficult to see
in the gauge fixed form of the equations.
\vskip 9pt
\noindent{\bf 5. The Interaction Term}
\vskip 9pt

So far, we have  considered only the linearized equations. Our task in this
section is to express the non-linear interaction term in eq.(4)
 in the language of states. For this purpose, it is very convenient to
use a coherent state representation for the states:
\bq
&&|F \rangle=\int Du\, Dv\, F_{m}(u,v) \exp\left(
\sum_{1}^{\infty}\frac{1}{k}\alpha_{-m}\beta_{-m}\right)\nonumber\\
&&\times\exp\left(iu_{0}.x+iv_{0}.P+ \sum_{1}^{\infty}(u_{m}\alpha_{-m}
+v_{m}\beta_{-m})\right) |0\rangle.
\eq
The states are constructed from the operators $\alpha_{m}$, $\beta_{m}$
and $P$ described in section (1), but they are now labeled by
$F(u,v)$,  a functional of variables $u^{\mu}_{m}$ and
$v^{\mu}_{m}$. Let the states representing $K_{m}$ and $K_{n}$ be labeled by
$F_{m}$ and $F_{n}$, and the state representing $[K_{m},K_{n}]$ by
$F_{(m,n)}$. The commutator $[K_{m},K_{n}]$ is easily calculated using
the coherent state representation. The result is
\bq
F_{(m,n)}(u,v)
=\int Du' Dv' \int Du'' Dv'' F_{m}(u',v') F_{n}(u'',v'')
 \prod_{i}\delta(u_{i}'+u_{i}''-u_{i})&&\nonumber\\
 \prod_{j}\delta(v_{j}'+v_{j}''-v_{j})
\left(\exp\left(iv_{0}'.u_{0}''+\sum_{1}^{\infty} k v_{k}'.u_{k}''\right)
-\exp\left(v_{0}''.u_{0}'+\sum_{1}^{\infty}k v_{k}''.u_{k}'\right)\right)&&.
\eq
It is convenient to define a star product [11] of the states  labeled by
$F_{m}$ and $F_{n}$ to be the state labeled by $F_{(m,n)}$,
$$
|F_{m}\rangle \star |F_{n}\rangle =|F_{(m,n)}\rangle,
$$
where $F_{(m,n)}$ is given by eq.(46). The star product defined in this
way is non-commutative but associative. Eq.(4) can now be cast into
the form
\be
L^{(0)}_{m}|F_{n}\rangle -L_{n}^{(0)}|F_{m}\rangle
-(m-n) |F_{m}\rangle \star |F_{n}\rangle =0.
\ee
So far, we have considered the open string. However, it is straightforward to
generalize everything we have done to the closed string by simply doubling
the number of operators. 

We have seen that the Virasoro algebra naturally gives rise to an
interaction term. The next question is whether it agrees with the
standard string interaction. It seems quite plausible that the standard 
string interaction should provide at least one particular solution to 
our equations, since it is highly unlikely that standard string
theory is eliminated by imposing Virasoro invariance. Of course,
the existence of additional solutions cannot a priori be ruled out.
One way to attack this problem is to gauge fix to the light cone gauge
and to compute the interaction term in that gauge. We hope to return 
to this problem in the future. 
\vskip 9pt
\noindent{\bf 6. The BRST Formulation}
\vskip 9pt

An elegant and convenient way of handling a theory invariant under a gauge
symmetry is the BRST method, which has been extensively used in the various
formulations of the string theory [11]. Here, we will first show how to
reformulate the linear equations for the open string (eq.(16)) as the
nilpotency condition for the BRST operator, and then we will generalize
to include the interaction term.
 The standard 
  BRST operator for the free Virasoro algebra is given by,
\be
Q= \sum_{-\infty}^{\infty}L^{(0)}_{m}c_{-m}-\frac{1}{2}\sum_{-\infty}
^{\infty}(m-n) c_{-m} c_{-n} b_{m+n},
\ee
where $L^{(0)}_{m}$ is given by eq.(2), and $b$'s and $c$'s satisfy
 the usual anticommutation relations:
$$
\{c_{m},b_{n}\}=\delta_{m,-n}.
$$
We know that $Q$ is nilpotent at the critical dimension:
$$
Q^{2}=0.
$$
Next, we wish to determine the form of $Q$ in the state representation.
 This can
be done directly by essentially going through the steps that led to
eq.(14). In the present case, we start with 
  normal ordered operator products
 constructed out of b's and c's in addition to the $\alpha$'s, and
represent these operators by states as in section (2). To do this, we
have to introduce the operators $\beta$ as before, and also we need
two sets of addditional anticommuting operators, $\bar{b}_{m}$ and
$\bar{c}_{m}$, which are ``partners'' of the $b$'s and the $c$'s. Next we
compute the anticommutator of $Q$ with the operator product.
 This anticommutator induces a mapping from the initial set of 
states to the set of states representing the anticommutator, which
then provides the state representation for $Q$. This procedure, although
straightforward, gives a complicated and unwieldy result .
Instead, we will try and guess a simple and economical form for $Q$, and
afterwards check that the equation
\be
Q|s\rangle=0
\ee
correctly reproduces eq.(16).

Our guess for the state representation for
 $Q$  is the following: We will keep eq.(48)
as it stands, but the $L^{(0)}_{m}$'s of eq.(2) will be replaced by those of
eq.(14). Also,
 our choice of the vacuum will be different from the usual one.
The standard vacuum is defined so that it is annihilated by all
 $c_{m}$ and $b_{m}$ for $m\geq 1$. Instead, our vacuum will satisfy
$$
b_{m}|0\rangle=0,
$$
for all $m$, and no other conditions. In particular,
 none of the $c$'s annihilate the vacuum. It is important to 
check whether  the nilpotency condition
$$
Q^{2}=0
$$
is satisfied. 
 Generically, this condition is violated by the central
charge of the Virasoro algebra. Only in the critical dimension, the
central charge of the Virasoro generators for the matter fields cancel
against the central charge of the Virasoro generators for the ghost
fields. In the present  case, the generators for the matter fields, given by
eq.(14), have zero central charge. On the other hand, it is easy to
verify that, with our new definition of the vacuum, the Virasoro generators
of the ghost sector also have no central charge, and the nilpotency
condition is still satisfied.

 We will now briefly discuss 
 equation (49).
At ghost number zero, there are no solutions to this equation; in 
particular, the vacuum is no solution. At ghost number 1, the state
\be
|s\rangle= \sum_{-\infty}^{\infty}c_{k}|k\rangle,
\ee
is a solution if and only if the states $|k\rangle$ satisfy eq.(16).
As usual, the gauge transformations (18) are generated by the BRST
transformations
$$
|s\rangle\rightarrow |s\rangle + Q|g\rangle,
$$
where $|g\rangle$ has ghost number zero. It is now easy to incorporate
the interaction term; identifying the states $|k\rangle$ in eq.(50) with
the coherent states of eq.(45),
$$
|s\rangle\rightarrow \sum c_{-k} |F_{k}\rangle,
$$
 we define a star product of two states $|s\rangle$ and $|s'\rangle$ by
$$
|s\rangle \star |s'\rangle=\sum c_{-k}\, c_{-l}\,|F_{k}\rangle \star
|F'_{l}\rangle.
$$
The equations with interaction(eq.(47)) can then be written as
\be
Q|s\rangle + |s\rangle \star |s\rangle=0.
\ee
It is easy to see that this equation is consistent with the nilpotency of
$Q$. After all, all we have done is to rewrite eq.(4) in a fancier language,
and $Q^{2}=0$ is equivalent to the Jacobi identity for the $L$'s.

The BRST formalism makes
it  possible to simplify and streamline the derivation
 results of the earlier sections, although we will not try to do it
here. We should also point out that
 there are  some problems with the unconventional
definition of the vacuum state we have adopted; for example, the space
of states defined in this fashion is not self dual and as a result,
the BRST operator is not self adjoint. The alternative construction
that we have sketched following eq.(48) avoids this problem, but as we
have already mentioned, leads to a complicated expression.

As a final comment, we would like to point out that so far, we have
only considered the equations of motion for the string states.
However, given  the BRST operator and the star product, 
 it is not too difficult to
construct an action from which these equations  follow. For example,
proceeding as in reference [11], one could try the action
\be
I=\langle s|Q|s\rangle + \langle s|\left(|s\rangle \star
|s\rangle\right),
\ee
where $|s\rangle$ is now a general state of odd ghost number. We hope to
return to this problem in the future.

\vskip 9pt
\noindent{\bf 7. Conclusions and Future Directions}

The basic idea of this paper was to derive string dynamics from the
realization of the conformal algebra on a suitable operator space.
The operator space we have chosen is the familiar one that results
from quantizing the flat space free field background. By requiring
the closure of the conformal algebra, we have derived a set of 
dynamical equations for the string. The novelty of this approach lies
in the identification of the non-commuting operators with string
coordinates. This results in the enlargement of the coordinate space;
but in compensation, there is invariance under
 a large gauge group given by eq.(6). To reach the standard picture of the
string, one has to do some suitable gauge fixing.

There are many interesting questions left to answer. For example, can
one gain some insight into the non-pertubative symmetries of string
theory by exploiting the new gauge invariance? Some other new avenues
of exploration are the introduction of fermions and supersymmetry,
and non-trivial backgrounds corresponding to various branes. 
\vskip 9pt

\newpage
{\bf References}
\begin{enumerate}
\item J.Polchinski, \emph{String Theory}, Cambridge University Press,
1998,\\
M.B.Green, J.H.Schwarz and E.Witten,\emph{Superstring Theory},
Cambridge University Press, 1987.
\item For a review, see:
A.Sen, \emph{An Introduction to Non-Perturbative String Theory},
hep-th/9802051,
M.Haack, B.K\"{o}rs and D.L\"{u}st, \emph{Recent Developments in
String Theory: From Perturbative Developments to M-Theory},
hep-th/9904033.
\item J.Dai, R.G.Leigh and J.Polchinski, Mod. Phys. Lett. {\bf A4},
(1989) 2073.
\item W.Taylor IV., \emph{Lectures on D-Branes, Gauge Theory and
M(atrices)},hep-th/9801182.
\item J.Maldacena, Adv.Theor. Math. Phys.{\bf 2}(1998) 231.
\item For a review see:
O.Aharony, S.S.Gubser, J.Maldacena, H.Ooguri and Y.Oz,
\emph{Large N Field Theories, String Theory and Gravity},
hep-th/9905111.
\item T.Banks, W.Fischler, S.Shenker and L.Susskind, Phys.Rev.
{\bf D55}(1997) 5112.
\item D.Bigatti and L.Susskind, \emph{Review of Matrix Theory},
hep-th/9712072
\item C.B.Thorn, Phys.Rep.{\bf 175}(1989)1.
\item M.Bochicchio, Phys. Lett. {\bf B193}(1987) 31.
\item E.Witten, Nuc. Phys.{\bf B 268}(1986) 253.
\item B.Zwiebach, Nuc. Phys.{\bf B 480}(1996) 541, A.Sen and
B.Zwiebach, Nuc. Phys.{\bf B 423}(1994) 580.
\item M.Kaku and K.Kikkawa, Phys. Rev.{\bf D 10}(1974) 1110.
\item W.Taylor IV, Phys. Lett.{\bf B 394}(1997) 283.
\item R.Dijkgraaf, E.Verlinde and H.Verlinde, Nuc. Phys.{\bf B 500}
(1997) 43.
\item L.Alvarez-Gaume, D.Z.Freedman and S.Mukhi, Ann. Phys.{\bf 134}
(1981) 85.
\item D.Friedan, Phys. Rev. Lett. {\bf 45}(1980) 1057.
\item D.Friedan, E.Martinec and S.Shenker, Nuc. Phys.{\bf B 271}
(1986) 93.
\item  C.Lovelace, Phys. Lett.{\bf B 135}(1984) 75.
\item C.G.Callen, D.Friedan, E.Martinec and M.J.Perry, Nuc. Phys.
{\bf B 262}(1985) 593.
\item T.Banks and E.Martinec, Nuc. Phys. {\bf B 294}(1987) 733.
\item J.Hughes, J.Liu and J.Polchinski, Nuc. Phys. {\bf B 316}
(1989) 15.
\item K.Bardakci and L.M.Bernardo, Nuc. Phys.{\bf B 505}(1997) 463;
K.Bardakci, Nuc. Phys.{\bf B 524}(1998) 545.
\item A.Giveon, M.Porrati and E.Rabinovici, Phys. Rep.{\bf 244}
(1994) 77.
\item M.R.Douglas and C.Hull, J. High Energy Phys. {\bf 02}(1998) 008.
\item A.Connes, M.R.Douglas and A.Schwarz, J. High Energy Phys.{\bf 02}
(1998) 003.
\item M.Rieffel and A.Schwarz, \emph{Morita Equivalence of
Multidimensional Noncommutative Tori}, math.QA/98033057.
\item D.Brace, B.Morariu and B.Zumino, \emph{Dualities of the Matrix
Model from T-Duality of the Type II String}, hep-th/9810099
\item P.-M.Ho, and Y.-S.Wu,\emph{Noncommutative Gauge Theories in
Matrix Theory}, hep-th/9801147.
\item C.Hofman and E.Verlinde, \emph{Gauge Bundles and Born-Infeld
on the Noncommutative  Torus}, hep-th/9810219.
\item F.Ardalan, H.Arfei and M.M.Sheikh-Jabbari, \emph{Mixed Branes
and M(atrix) Theory on Noncommutative Torus}, hep-th/9803067.
\end{enumerate}
\end{document}